\documentclass[12pt]{JHEP3}

\usepackage{latexsym,amsfonts,amsmath,amssymb,wasysym,stmaryrd,enumerate,epsfig}
\usepackage{graphicx}
\usepackage{amsmath}
\usepackage{amsfonts}
\usepackage{amssymb}
\usepackage{theorem}

\def\beq{\begin{equation}}
\def\eeq{\end{equation}}
\def\bea{\begin{eqnarray}}
\def\eea{\end{eqnarray}}
\def\bean{\begin{eqnarray*}}
\def\eean{\end{eqnarray*}}
\def\al{\alpha}
\def\ka{\kappa}
\def\E{\textbf{E}}
\def\C{\textbf{C}}
\def\K{\textbf{K}}
\def\ud{\textrm{d}}

\title{Einstein-Gauss-Bonnet metrics: black holes, black strings and a staticity theorem.}

\author{{\bf C. Bogdanos}$^{1}$, {\bf C. Charmousis}$^{1,2}$, {\bf B. Gout\'eraux}$^{1}$ and {\bf R. Zegers}$^{3}$\\
$^{1}${\it \href{http://www.th.u-psud.fr/}{Laboratoire de Physique Th\'eorique
CNRS UMR 8627, \href{http://www.u-psud.fr}Universit\'e Paris-Sud 11 91405 Orsay Cedex, France}}\\
$^{2}${{ \it Laboratoire de Math\'ematiques et Physique Th\'eorique (LMPT) CNRS  UMR 6083 Universit\'e Fran\c cois Rabelais - Tours, France}}\\
$^{3}${\it \href{http://www.cpt.dur.ac.uk/}{Centre for Particle Theory, \href{http://maths.dur.ac.uk/}Department of Mathematical Sciences,
\href{http://www.dur.ac.uk/}Durham University, South Road, Durham DH1 3LE, U.K.}}}

\date{today}
\preprint{LPT-09-46}
\abstract{We find the general solution of the 6-dimensional Einstein-Gauss-Bonnet equations in a large class of space and time-dependent warped geometries. Several distinct families of solutions are found, some of which include black string metrics, space and time-dependent solutions and black holes with exotic horizons. Among these, some are shown to verify a Birkhoff type staticity theorem, although here, the usual assumption of maximal symmetry on the horizon is relaxed, allowing exotic horizon geometries. We provide explicit examples of such static exotic black holes, including ones whose horizon geometry is that of a Bergman space. We find that the situation is very different from higher-dimensional general relativity, where Einstein spaces are admissible black hole horizons and the associated black hole potential is not even affected.
In Einstein-Gauss-Bonnet theory, on the contrary, the non-trivial Weyl tensor of such exotic horizons is exposed to the bulk dynamics through the higher order Gauss-Bonnet term, severely constraining the allowed horizon geometries and adding a novel charge-like parameter to the black hole potential. The latter is related to the Euler characteristic of the four-dimensional horizon and provides, in some cases, additional black hole horizons.}

\begin{document}

\section{Introduction}
Gravitational theories in more than four spacetime dimensions have gained a lot of attention over the past three decades. Although these ideas go back to the early days of General Relativity, with the introduction of Kaluza-Klein theories\cite{Kaluza:1921tu,Klein:1926tv}, it was the advent of String Theory that revived the notion of higher-dimensional spacetimes as not just an interesting theoretical possibility, but as a necessary ingredient of a unified picture of elementary interactions. Not surprisingly, the mere extension of General Relativity by considering extra spacelike dimensions can immediately lead to very non-trivial alterations in the theory. The inclusion of additional structure in the gravitational action, such as Gauss-Bonnet and Lovelock\cite{Lovelock:1971yv} terms, or brane-like components\cite{Akama:1982jy,Rubakov:1983bb,Antoniadis:1990ew,ArkaniHamed:1998rs,Antoniadis:1998ig,Randall:1999ee,Randall:1999vf} increases even further the diversity of the models available and gives rise to a rich phenomenology, one which is actively investigated these days. The long standing problems in gravity, such as gravitational collapse, the initial singularity conditions, a number of open cosmological problems such as dark matter and accelerated expansion of the universe, as well as the elusive quantum theory have accumulated over the years to a general consensus which casts considerable doubt on General Relativity as the final word on gravity in a number of different regimes. This acts as a further motivation to give extra-dimensional theories serious consideration as possible routes to a more complete description of this fundamental interaction.

Gauss-Bonnet extensions of General Relativity (GR) have been motivated from a string-theoretical point of view as a version of higher-dimensional gravity, since this sort of modification also appears in low energy effective actions in this context \cite{Zwiebach:1985uq} (see also the points raised in \cite{Gross:1986mw}). The same gravitational term is also present in the case of Lovelock theory (for recent reviews see \cite{Deruelle:2003ck}, \cite{Charmousis:2008kc}), which provides a unique and unambiguous classical extension of GR in arbitrary dimensions. The theory of such extended gravity theories has been extensively studied (see for example \cite{Boulware:1985wk,Wheeler:1985nh,Wheeler:1985qd,Deser:2002jk,Myers:1988ze,Dehghani:2005zm,Maeda:2006iw,Maeda:2006hj,Kofinas:2007ns,Kofinas:2006hr}), especially in conjunction with braneworlds (see for example \cite{Gregory:2003px,Padilla:2004mc, Charmousis:2005ez,Charmousis:2005ey,Papantonopoulos:2007fk,Charmousis:2008bt}). Studies of the cosmology of these setups have also provided insight into the possible relevance of the Gauss-Bonnet gravitational term to 4-dimensional inflation and the accelerated cosmic expansion (see for example \cite{Amendola:2005cr,Koivisto:2006ai,Amendola:2008vd,Amendola:2007ni,Koivisto:2006xf}).

It is well-known that Birkhoff's theorem, when considered in the context of higher-dimensional GR ($n>4$), remains valid and is in fact amplified in terms of its generality\cite{Gibbons:2002pq,Gibbons:2002th}.  The original Birkhoff theorem states that, in four dimensions, any spherically symmetric solution to Einstein's equations in the vacuum is necessarily locally static, a very important result with many applications when considering the gravitational field of ordinary stars. It is worth mentioning that, in four dimensions, there also exists a form of reciprocal to Birkhoff's theorem. First, the horizon of an asymptotically flat stationary black hole must have the topology of a 2-sphere \cite{Hawking:1971vc}. Moreover, under quite general assumptions, Israel's theorem states that every static black hole whose horizon has the topology of a 2-sphere is isometric to the Schwarzschild solution \cite{PhysRev.164.1776,Israel:1967za}. In other words, not only is its horizon topologically a 2-sphere but it also has the metric of the \emph{round} 2-sphere. In higher dimensions, these well established four-dimensional uniqueness results just fail : on one hand, because the topology of the horizon is less restricted \cite{Cai:2001su, Galloway:2005mf, Helfgott:2005jn}; on the other hand, because, even if one insists on having a particular horizon topology, the actual geometry on this horizon is much less constrained. This leaves room for
Birkhoff's theorem to remain valid not only for a constant curvature horizon, but also for horizons which belong to the more general class of Einstein spaces. Substituting the usual $(n-2)$-sphere of the horizon geometry (in the case of an $n$-dimensional spacetime) with an $(n-2)$-dimensional Einstein manifold will not alter the black hole potential and the previous solution remains valid and static. Spherical symmetry is no longer a prerequisite for staticity. The structure of the space transverse to the horizon is in this way not affected by the details of the internal geometry, as long as the latter continues to be an Einstein space.  Such exotic black holes are accompanied by classical instabilities  \cite{Gibbons:2002pq,Gibbons:2002th} similar to those of the black string \cite{Gregory:1993vy}. In fact black string metrics can be Wick rotated to a subclass of metrics with exotic horizons. The exotic horizon is nothing but the Euclidean version of 4 dimensional Schwarzschild. Therefore one could entertain the possibility that the additional unphysical exotic black holes are just an artifact of not considering the full classical gravity theory in higher dimensions. In fact it was shown by Lovelock in the early 70's \cite{Lovelock:1971yv} that in higher than 4 dimensions specific higher order gravity terms have to be added to the usual Einstein Hilbert action in order to preserve the unique properties of general relativity in 4 dimensions (for a discussion and the geometric properties see \cite{Charmousis:2008bt}). These higher order gravity terms, which include the Ricci and Gauss-Bonnet scalar are dimensionally extended Euler Poincar\'e densities of $2,4$ dimensional and so forth manifolds.

In fact the situation is very different when higher order curvature terms such as the Gauss-Bonnet term, are introduced. As was recently shown in \cite{Dotti:2005rc}, the presence of the Gauss-Bonnet term can be quite restrictive for the geometry of the horizon of a black hole, compared to ordinary GR results (see also \cite{Barcelo:2002wz} ). Intuitively, this can be understood as follows: in GR, Einstein's equations only involve the Ricci tensor, whereas the Einstein-Gauss-Bonnet field equations expose the entire Riemann curvature tensor to the dynamics.
In \cite{Dotti:2005rc}, the authors considered a static spacetime with generic Einstein space as an $n-2$ dimensional subspace and then analysed the field equations. They found that the rank two tensor $C^{acde}C_{bcde}$, where $C_{abcd}$ is the Weyl tensor, is representative of the new solutions and only horizons satisfying the appropriate conditions on $C^{acde}C_{bcde}$ are allowed.

In this paper, we investigate an extension of Birkhoff's theorem to the six-dimensional Einstein-Gauss-Bonnet theory{\footnote{Birkhoff's theorem for Einstein-Gauss-Bonnet theorem was demonstrated by Wiltshire \cite{Wiltshire:1985us}. Here, when refering to this theorem we will be using the slightly generalised version discussed in \cite{Charmousis:2002rc}. }}, allowing arbitrary $4$-dimensional horizon geometries and of course time dependence in the metric. In particular, we show that Birkhoff's theorem holds quite generically though the theory is far more complex. Although the allowed horizon geometries are far more restricted than in dimensionally extended GR, in agreement with \cite{Dotti:2005rc}, we shall see that they need not be maximally symmetric. Namely, it will suffice that they be Einstein spaces and that the invariant built out by squaring their Weyl tensor be a constant.

We would like to stress that the 6-dimensional case is very special: in 5 dimensions, the Weyl tensor is identically zero, whereas in more than 6 dimensions, Lovelock theory dictates the presence of a higher order gravity term in the action. Furthermore, in 6 dimensions the 4-dimensional horizon geometry allows for a non trivial 4-dimensional Gauss-Bonnet term which when integrated over the horizon surface gives a topological charge, the 4-dimensional Euler-Poincar\'e characteristic.

The paper is organized as follows. We first derive the general Einstein-Gauss-Bonnet field equations for the class of metrics considered throughout the paper. We then systematically solve these equations. Just as in the Lovelock extension of Birkhoff's theorem \cite{Zegers:2005vx}, we encounter two distinct classes of solutions, plus a third particular one
(see also \cite{Dotti:2008pp} for the classification of the static metrics). The first of them comes along with a fine-tuning of the parameters of the theory, which corresponds in our case to the Born-Infeld limit, and leads to an underdetermined system of equations. The solutions of this branch are not necessarily static. From the second branch we obtain a set of static solutions including black hole solutions, where the horizon is an Einstein space of constant Ricci scalar and constant $C^{acde}C_{bcde}$, and generalizations of the Nariai solution. We also encounter a branch of solutions obeying the staticity theorem but with non-Einstein space horizons. The third class of solutions is unwarped, and contains both fine-tuned and non-fine-tuned solutions, some of them static, with or without Einstein horizon. We then present a number of explicit examples of such horizon manifolds, for instance products of $2$-spheres and the Bergman metric, as well as horizons with a possible relevance for codimension two braneworlds.

\section{Action and Conventions}
We begin by considering the Einstein-Gauss-Bonnet action with a cosmological constant in six dimensions
\beq
S^{\left( 6 \right)}  = \frac{{{M^{(6)}}^4 }}{2}\int {d^6 x\sqrt { - g^{\left( 6 \right)} } \left[ {R - 2\Lambda  + \alpha \hat G} \right]} \,,
\eeq
where $M^{(6)}$ is the fundamental mass scale in six-dimensional spacetime,  $\hat G$ the Gauss-Bonnet density defined as
\beq
\hat G = R_{ABCD} R^{ADCB}  - 4R_{AB} R^{AB}  + R^2 \, ,
\eeq
and $\Lambda$ the cosmological constant. Using these conventions we can vary the action with respect to the metric to derive the field equations
\beq
{\cal E}_{AB}  = G_{AB}  + \Lambda g_{AB}  + \alpha H_{AB}  = 0\,,
\label{gbeq}
\eeq
where $G_{AB}$ stands for the Einstein tensor. Uppercase indices refer to six-dimensional coordinates. We have also introduced the Lanczos or Gauss-Bonnet tensor,
\beq
H_{AB}  = \frac{{g_{AB} }}{2}\hat G - 2RR_{AB}  + 4R_{AC} R^C _{\;B}  + 4R_{CD} R^{C\;D} _{\;A\;B}  - 2R_{ACDE} R_B ^{\;CDE} \,.
\eeq
Interestingly, the latter can also be written using the following rank four tensor
\beq P_{ABCD} \doteq R_{ABCD}+R_{BC}g_{AD}-R_{BD}g_{AC}-R_{AC}g_{BD}+R_{AD}g_{BC}+\frac{1}{2} R g_{AC} g_{BD}-\frac{1}{2} R g_{BC} g_{AD} ,\eeq
as
\beq
\label{ddual}
H_{AB}=P_{ACDE}R_B{ }^{CDE}-\frac{{g_{AB} }}{2}\hat G \, .
\eeq
The tensor $P_{ABCD}$ has several interesting properties: it is divergence free since the Bianchi identities of the curvature tensor are simply $\nabla^D P_{ABCD}=0$. It has also has the same index symmetries as the Riemann curvature tensor. Tracing two of its indices yields $P^{B}{}_{ACB}=G_{AC}$, which in turn yields the divergence free property of the Einstein tensor. In rather loose terms, one can say that $P$ is the curvature tensor associated to the Einstein tensor, just as the Ricci tensor is associated to the Riemann tensor. In four dimensions, this statement is far more precise since $P_{ABCD}$ coincides with the double dual (i.e. for each pair of indices) of the Riemann tensor $^\star R^{CD}{}_{AB}^\star \doteq -\frac{1}{2} \epsilon^{ABMN}\,R_{MN}{ }^{RS} \, \frac{1}{2} \epsilon_{RSCD}$, where $\epsilon_{ABCD}$ is the rank 4 Levi-Civita tensor. In 4 dimensions we have $H_{AB}=0$ thus picking up the following Lovelock identity (for extensions see \cite{Edgar:2001vv}),
\beq
\label{lock}
P_{ACDE}R_B{ }^{CDE}=\frac{{g_{AB} }}{2}\hat G
\eeq
which will be useful to us later on.

In order to proceed with the solution of the equations, we are now going to choose an appropriate symmetry for the metric. We distinguish between the transverse 2-space, which also carries the timelike coordinate $t$, and the internal 4-space, which is going to represent the possible horizon line element of the six-dimensional black hole. The metric of the internal space $h_{\mu\nu}$ is an arbitrary metric of the internal coordinates $x^\mu, \mu=0,1,2,3$ but we are imposing that the internal and transerse spaces are orthogonal to each other. This is an additional hypothesis we have to make since $h_{\mu\nu}$ is not a homogeneous metric and because our six-dimensional space is {\it not} an Einstein space (in GR such an orthogonal foliation is possible for an Einstein metric). At a loss of a better name we will call this a warped metric Ansatz. Guided by the analogous procedure of analyzing Birkhoff's theorem we write the metric as
\beq
\label{metric}
ds^2  = e^{2\nu \left( {t,z} \right)} B\left( {t,z} \right)^{ - 3/4} \left( { - dt^2  + dz^2 } \right) + B\left( {t,z} \right)^{1/2} h^{\left( 4 \right)} _{\mu \nu } \left( x \right)dx^\mu  dx^\nu  \,.
\eeq
Lowercase greek indices correspond to internal coordinates of the 4-space. We then switch the coordinates of the transverse space to light-cone coordinates,
\beq
u = \frac{{t - z}}{{\sqrt 2 }},\quad v = \frac{{t + z}}{{\sqrt 2 }}.
\eeq
in terms of which the metric reads
\beq
ds^2  =  - 2e^{2\nu \left( {u,v} \right)} B\left( {u,v} \right)^{ - 3/4} dudv + B\left( {u,v} \right)^{1/2} h^{\left( 4 \right)} _{\mu \nu } \left( x \right)dx^\mu  dx^\nu  \,.
\eeq
Using the above prescription, we are now able to write down the equations of motion. The $u u $ and $\upsilon \upsilon $ equations yield
\beq
\label{uu}
{\cal E}_{uu}=\frac{2 \nu_{,u} B_{,u}- B_{,uu}}{B} \left[ 1+\alpha \left( B^{-1/2} R^{(4)}+\frac{3}{2} e^{-2\nu} B^{-5/4} B_{,u} B_{,v}  \right) \right]\,,
\eeq
\beq
\label{vv}
{\cal E}_{vv}=\frac{2 \nu_{,v} B_{,v}- B_{,vv}}{B} \left[ 1+\alpha \left( B^{-1/2} R^{(4)}+\frac{3}{2} e^{-2\nu} B^{-5/4} B_{,u} B_{,v}  \right) \right].
\eeq
The off-diagonal equation reads
\bea
\label{uv}
{\cal E}_{uv}  &=&\frac{{B_{,uv} }}{B} - \Lambda e^{2\nu } B^{ - 3/4}  + \frac{\alpha }{2}e^{2\nu } B^{ - 7/4} \hat G^{(4)} \nonumber\\
&+& R^{(4)} \left[ {\frac{1}{2}e^{2\nu } B^{ - 5/4}  - \alpha B^{ - 3/2} \left( {\frac{1}{2}\frac{{B_{,u} B_{,v} }}{B} - B_{,uv} } \right)} \right] \nonumber\\
&+&  \alpha e^{ - 2\nu } B^{ - 5/4} \left[ { - \frac{{15}}{{16}}\left( {\frac{{B_{,u} B_{,v} }}{B}} \right)^2  + \frac{3}{2}\frac{{B_{,u} B_{,v} }}{B}B_{,uv} } \right] \,.
\eea
We also have the $\mu \nu$ equations, which can be brought into the form
\bea
\label{munu}
{\cal E}_{\mu \nu}&=& G_{\mu \nu }^{(4)}  - e^{ - 2\nu } B^{1/4} \left( {\frac{3}{4}B_{,uv}  + 2B\nu _{,uv} } \right)h^{\left( 4 \right)} _{\mu \nu }  + \Lambda B^{1/2} h^{\left( 4 \right)} _{\mu \nu } \nonumber\\
&+& \frac{3}{2}\alpha e^{ - 4\nu } \left( {B_{,uu}  - 2\nu _{,u} B_{,u} } \right)\left( {B_{,vv}  - 2\nu _{,v} B_{,v} } \right)h^{\left( 4 \right)} _{\mu \nu } \nonumber \\
&-&\alpha e^{ - 4\nu } \left[ {\frac{{45}}{{32}}\left( {\frac{{B_{,u} B_{,v} }}{B}} \right)^2  - \frac{{21}}{8}\frac{{B_{,u} B_{,v} }}{B}B_{,uv}  + \frac{3}{2}B_{,uv} ^2  + 3B_{,u} B_{,v} \nu _{,uv} } \right]h^{\left( 4 \right)} _{\mu \nu }\nonumber\\
&-&
\alpha e^{ - 2\nu } B^{ - 1/4} \left( {\frac{3}{4}\frac{{B_{,u} B_{,v} }}{B} - \frac{1}{2}B_{,uv}  + 4B\nu _{,uv} } \right)\left( {R^{(4)} h^{\left( 4 \right)} _{\mu \nu }  - 2R_{\mu \nu }^{(4)} } \right)
\,\,.
\eea

In this way, we have decomposed the gravitational equations into expressions depending on either transverse space quantities, or internal coordinates. The integrability conditions, \cite{Bowcock:2000cq}, are unchanged compared to the original version of the theorem \cite{Charmousis:2002rc}, and this will permit us to obtain the staticity conditions.  Furthermore, the internal geometry of the horizon only enters these equations through expressions involving the four-dimensional Gauss-Bonnet scalar density, the Ricci tensor and scalar of the internal metric $h_{\mu\nu}$. Note the absence of $H^{(4)}_{\mu\nu}$ terms due to the fact that internal space is 4-dimensional. Note also that terms proportional to the Gauss-Bonnet coupling constant are the ones responsible for the appearance of $R^{(4)}$ and $R^{(4)}_{\mu \nu}$ and in this way, the Gauss-Bonnet term exposes the internal geometry to the transverse space dynamics in a non-trivial way, something which would obviously not occur in ordinary General Relativity. As we will see, this decomposition imposes severe constraints on the allowed form of the horizon geometry in order to get a spacetime solution.

\section{Exact Solutions and Staticity}

The $uu$ and $vv$ equations (\ref{uu}), (\ref{vv}) can lead to three different classes of solutions, depending on wether the first or second factor is zero (an additional class will emerge for constant $B$). The corresponding solutions have distinct characteristics and are thus treated separately in what follows. Class I and II are both warped solutions whereas for Class III we have $B=const.$.

\subsection{Class-I}

This class corresponds to solutions which can have, in general, time dependence and, hence, for which a Birkhoff-type theorem does not hold. As we shall soon see, all of them imply $5 + 12\alpha \Lambda  = 0$. The latter corresponds to the so-called \emph{Born-Infeld} limit, an even-dimensional counterpart of the well-known odd-dimensional Chern-Simons limit in which the Lovelock action can be written as a Chern-Simons action for some (a)dS connection -- see e.g. \cite{Zanelli:2005sa}. In the Born-Infeld limit, the Lovelock action can be written as a Born-Infeld action for some curvature 2-form, hence its name. For the class of space-time metrics under consideration here, it typically leads to an underdetermined set of equations and the unconstrained components of the metric subsequently allow for a possible time-dependence. This is reminiscent of class-I Lovelock solutions with spherical, hyperbolic or planar symmetry \cite{Charmousis:2002rc, Zegers:2005vx} and is expectedly related to perturbative strong coupling problems as in the case of Chern-Simons gravity \cite{CharmousisPadilla2008}.

Setting the second factor of the $(uu)$ and $(vv)$ equations (\ref{uu}) equal to zero leads to the common equation
\beq
\label{classIcond}
1 + \alpha {B^{ - 1/2} R^{(4)}  + \frac{3}{2} \alpha e^{ - 2\nu } B^{ - 5/4} B_{,u} B_{,v} }  = 0\,,
\eeq
from which we can solve for the function $\nu(u,v)$ in terms of $B(u,v)$, according to
\beq
\label{int11}
\nu \left( {u,v} \right) = \frac{1}{2}\ln \left( { - \frac{{3\alpha }}{2}\frac{{B_{,u} B_{,v} }}{{B^{5/4} \left( {1 + \alpha B^{ - 1/2} R^{\left( 4 \right)} } \right)}}} \right) \,.
\eeq
Note that this equation immediately constrains the Ricci scalar $R^{(4)}$ of the internal space to be a constant. We are thus required to consider only horizon geometries of constant scalar curvature as candidate solutions. Substituting the above expression for $\nu(u,v)$ into (\ref{uv}) yields the two additional constraints,
\beq
5 + 12\alpha \Lambda  = 0,\,\,\hat G^{\left( 4 \right)}  = \frac{1}{6}{R^{\left( 4 \right)} }^2 \,.
\label{cons}
\eeq
The second of these tells us that the Gauss-Bonnet scalar $\hat G$ is also constant.
Taking the trace of (\ref{munu}) with $h_{\mu\nu}$ and performing the same substitution we end up with the equation
\beq
{\cal E} \equiv {\cal E}_\mu^{\mu} = \frac{{5 + 12\alpha \Lambda }}{{3\alpha }} = 0\,,
\eeq
Finally, we can rewrite the complete  equation (\ref{munu}) in terms of the trace as
\bea
{\cal E}_{\mu \nu} &=& \frac{1}{4}B^{1/2} {\cal E} h^{\left( 4 \right)} _{\mu \nu }  \nonumber\\
&+& \left( {R^{\left( 4 \right)} _{\mu \nu }  - \frac{1}{4}R^{\left( 4 \right)} h^{\left( 4 \right)} _{\mu \nu } } \right)\left[ {1 + 2\alpha e^{ - 2\nu } B^{ - 1/4} \left( {\frac{3}{4}\frac{{B_{,u} B_{,v} }}{B} - \frac{1}{2}B_{,uv}  + 4B\nu _{,uv} } \right)} \right].
\label{mneq}
\eea
Given the above mentioned constraints, the first term vanishes because it is proportional to ${\mathcal E}$. The second term can vanish in one of two ways giving us two distinct cases of Class-I solutions both verifying (\ref{int11}) and (\ref{cons}). We can either have
\beq
R^{\left( 4 \right)} _{\mu \nu }  = \frac{1}{4}R^{\left( 4 \right)} h^{\left( 4 \right)} _{\mu \nu } \,,
\eeq
which is the definition of a four-dimensional \emph{Einstein space} \footnote{In general, a d-dimensional Einstein space obeys $R_{\mu \nu }  = \frac{1}{d}Rh_{\mu \nu } $ where $R$ is a constant.}. Coupled with the condition  $\hat G^{\left( 4 \right)}  = \frac{1}{6}\left[R^{\left( 4 \right)} \right]^2$, this leads to
\beq
C^{\left( 4 \right)} _{\alpha \beta \mu \nu } C^{\left( 4 \right)\alpha \beta \mu \nu }  = 0\,,
\eeq
i.e. the square of the Weyl tensor of the internal space must be zero. We then have a \emph{constant curvature space}\footnote{A constant curvature space is defined by $R_{\mu \nu \rho \lambda}  = \frac{1}{d(d-1)}R\left(h_{\mu \rho }h{\nu\lambda}-h_{\mu\lambda}-h_{\nu\rho}\right) $ where $R$ is a constant.}.
Since (\ref{munu}) is in this way automatically satisfied, there is no dynamical equation defining the function $B(u,v)$ and thus the system of field equations becomes underdetermined. This is a typical feature of the Class-I solutions which have been discussed in \cite{Charmousis:2002rc}.

If, on the contrary, we demand the second factor in the second term of equation (\ref{mneq}) to be zero, the requirement for a four-dimensional Einstein space on the horizon of the black hole can be relaxed. Instead, we get a third order partial differential equation for $B(u,v)$, which reads
\bea
&&\left( {1 + \alpha B^{ - 1/2} R^{\left( 4 \right)} } \right)^2 \left( {B_{,u} ^2 B_{,vv} B_{,uv}  + B_{,v} ^2 B_{,uu} B_{,uv}  - B_{,u} ^2 B_{,v} B_{,uvv}  - B_{,v} ^2 B_{,u} B_{,uuv} } \right) \nonumber\\
&&+ \frac{{B_{,uv} }}{B}B_{,u} ^2 B_{,v} ^2 \left[ {\frac{3}{2} + \frac{5}{2}\alpha B^{ - 1/2} R^{\left( 4 \right)}  + \left( {\alpha B^{ - 1/2} R^{\left( 4 \right)} } \right)^2 } \right] \nonumber\\
&&- \frac{{B_{,u} ^3 B_{,v} ^3 }}{{B^2 }}\left[ {\frac{5}{4} + \frac{{17}}{8}\alpha B^{ - 1/2} R^{\left( 4 \right)}  + \frac{9}{8}\left( {\alpha B^{ - 1/2} R^{\left( 4 \right)} } \right)^2 } \right] = 0\,.\label{Beqn_class_notEinstein}
\eea
This equation can in principle be solved for $B(u,v)$, again for an internal space of constant Ricci scalar and given the constraints (\ref{cons}). Note that the horizon is not necessarily an Einstein space but instead we have the 4-dimensional geometrical constraint,
\beq
{C^{\left( 4 \right)}}^2+2{R^{\left( 4 \right)}_{\mu\nu}}^2=\frac12{R^{\left( 4 \right)}}^2= \mbox{constant}\,.
\eeq

We now summarize the results for the Class-I solutions. We distinguish two subclasses, both requiring the fine-tuning condition $5+12 \alpha \Lambda=0$, which is the six-dimensional version of the Born-Infeld gravity condition, and a constant Ricci scalar $R^{\left( 4 \right)}$ :
\begin{itemize}
	\item Class-Ia: we have an underdetermined system for the transverse dimension geometry (free function $B$ and (\ref{int11})) and an internal space which is an Einstein space of zero Weyl squared curvature, that is a \emph{constant curvature space},
	\item Class-Ib: A completely determined system of transverse dimensions (\ref{int11}), (\ref{Beqn_class_notEinstein}) with an internal geometry  obeying  (\ref{cons}) (non-zero Weyl curvature).
\end{itemize}
The former of the two subclasses is certainly incompatible with Birkhoff's theorem as demonstrated in \cite{Charmousis:2002rc}, whereas for the latter we could not find the general solution to \eqref{Beqn_class_notEinstein}.

\subsection{Class-II}
Class-II solutions are obtained by demanding, instead of (\ref{classIcond}), that
$$
\left\{\begin{array}{r}
		2 \nu_{,u} B_{,u}- B_{,uu} = 0\\
		2 \nu_{,v} B_{,v}- B_{,vv} = 0
      \end{array} \right.
	\label{nuBB}
$$
These integrability conditions are the same as in the case of ordinary GR. We will again assume that $B$ is not constant.

Equation (\ref{nuBB}) implies that
\beq e^{2\nu} = B_{,u} f(v) = B_{,v} g(u) \, , \label{enuBB}\eeq
for some functions $f$ and $g$, which, in turn, yields
$B=B(U+V)$, with $U=U(u)$ and $V=V(v)$. In this way, under the change of coordinates
\beq
\label{uvUV}
U = \frac{{\bar{z} - \bar{t}}}{{\sqrt 2 }}\,,\,\,\,V = \frac{{\bar{z} + \bar{t}}}{{\sqrt 2 }}\,,
\eeq
the function $B$ becomes independent of time and Birkhoff's theorem holds. Additionally, rewriting (\ref{enuBB}), $\nu(u,v)$ is now defined as
\beq
e^{2\nu }  = B'U'V'\,,
\eeq
where primes denote differentiation with respect to the single argument of each function. Under (\ref{uvUV}), we get $e^{2 \nu}= \partial_{\bar{z}} B$. The $uu$ and $vv$ equations thus determine the staticity of the metric, as well as the relation between $B$ and $\nu$. We can then determine $B(u,v)$, or equivalently the form of the black hole potential, from the $uv$ equation. Taking advantage of the already deduced staticity, we can express this as
\bea
B'' &+& \frac{1}{2}R^{\left( 4 \right)} B^{ - 1/4} B' - \frac{{15}}{{16}}\alpha B^{ - 9/4} B'^3  + \frac{3}{2}\alpha B^{ - 5/4} B''B'   \nonumber\\
&-& \frac{1}{2}\alpha R^{(4)} B^{ - 3/2} B'^2 + \alpha B^{ - 1/2} R^{\left( 4 \right)} B'' + \frac{1}{2}\alpha B^{ - 3/4} B'\hat G^{\left( 4 \right)}  - \Lambda B^{1/4} B' = 0\,.
\eea
Inspection of the above expression leads to the conclusion that a priori only solutions with a constant Ricci scalar and Gauss-Bonnet density for the internal space are permissible. However, this is not always the case, we have to be cautious of special cases. Upon integration, this leads to a quadratic equation for $B'$. We can then solve for $B'$ and determine the black hole potential $V$
\beq
\label{Class2a}
ds^2  =  - V\left( r \right)dt^2  + \frac{{dr^2 }}{{V\left( r \right)}} + r^2 h^{\left( 4 \right)} _{\mu \nu } \left( x \right)dx^\mu  dx^\nu \,,
\eeq
using the change of variables $r=B^{1/4}$. The corresponding potential turns out to be
\beq
V(r) = \frac{{R^{\left( 4 \right)} }}{{12}} + \frac{{r^2 }}{{12\alpha }}\left[ {1 \pm \sqrt {1 +\frac{12\alpha \Lambda}{5}  +\frac{{\alpha ^2 \left({R^{(4)}}^2-6\hat G^{(4)} \right)}}{{r^4 }} + 24\frac{{\alpha M}}{{r^5 }}} } \right]\,,
	\label{potentialclassII}
\eeq
where $M$ is an integration constant independent of $x$, related to the mass of the six-dimensional black hole \footnote{We note that the Gauss-Bonnet coupling constant has dimensions $mass^{-2}$, $k$ of $mass$ and $\kappa$ is dimensionless. The latter is justified by the fact that the internal metric $h^{(4)}_{\mu \nu} dx^{\mu} dx^{\nu}$ is multiplied by $r^{2}$, so the internal coordinates must be of an angular nature and carry no dimension. Consequently, derivatives with respect to them as well as the Riemmann, Ricci and Weyl tensor are dimensionless.}.

We now turn to the $\mu \nu$ equations (\ref{munu}). Taking the trace with respect to the internal metric  leads to the expression
\bean
{\cal E} &=& 4\Lambda  - R^{\left( 4 \right)} B^{ - 1/2}  - B^{ - 1/4} \left( {3\frac{{B''}}{{B'}} + 4\frac{{BB''}}{{B'^2 }} - 4\frac{{BB''^2 }}{{B'^3 }}} \right) \\
&-& \alpha B^{ - 1/2} \left( {\frac{{45}}{8}\frac{{B'^2 }}{{B^2 }} - \frac{{21}}{2}\frac{{B''}}{B} + 6\frac{{B'''}}{{B'}}} \right) \\
&-& \alpha R^{\left( 4 \right)} B^{ - 3/4} \left( {\frac{3}{2}\frac{{B'}}{B} - \frac{{B''}}{{B'}} + 4\frac{{BB'''}}{{B'^2 }} - 4\frac{{BB''^2 }}{{B'^3 }}} \right) = 0\,.
\eean
It can be shown that this equation can be rewritten as $ - \partial _v \left( {\frac{{B^{3/4} }}{{B'}}{\mathcal E}_{uv} } \right) = 0$, which is identically satisfied as a Bianchi identity.

The $\mu \nu$ equation then gives,
\beq
	0=\left(R^{\left( 4 \right)}_{\mu\nu}-\frac14R^{\left( 4 \right)}h_{\mu\nu}\right)\left[1 + \alpha B^{ - 1/4} \left( {\frac{3}{2}\frac{{B'}}{B} - \frac{{B''}}{B'} + 8\frac{{BB'''}}{{B'^2 }} - 8\frac{{BB''^2 }}{{B'^3 }}} \right) \right]
	\label{munueqclassII}
\eeq
Therefore, we have two distinct cases, depending on which of the two factors of \eqref{munueqclassII} cancels.

For the first case the horizon has to  be an Einstein space with constant scalar curvature, defined by $R^{\left( 4 \right)}_{\mu \nu}= 3 \kappa h_{\mu \nu}$. This is similar to ordinary GR. However given that $\hat G^{(4)}$ is also  constant we have that $C^{\alpha \beta \gamma \mu} C_{\alpha \beta \gamma \mu}=4\Theta $ where $\Theta$ is a positive constant. This is the solution obtained by \cite{Dotti:2005rc}. Now using the properties of the $P_{\mu\nu\alpha\beta}$ tensor and (\ref{lock}) we immediately get,
\beq
C^{\alpha \beta \gamma \mu} C_{\alpha \beta \gamma \nu}=\Theta \delta^\mu_\nu
\eeq
This is a supplementary condition imposed on the usual Einstein space condition for the horizon. Both have a similarity in that we ask for (part of) a curvature tensor to be analogous to the spacetime metric. The main difference being that the curvature tensor in question here is the Weyl tensor and, given its symmetries, it is actually its square which is analogous to the spacetime metric. Clearly horizons with $\Theta\neq 0$ will not be homogeneous spaces and not even asymptotically so in the non-compact cases. We will see in a forthcoming section that they can be related to squashed sphere geometries. Another interesting point is that the Gauss-Bonnet scalar, whose spacetime integral is the Euler characteristic of the horizon, has to be constant. In other words the Euler Poincar\'e characteristic of the horizon is in this case simply the volume integral of the horizon. In this sense $\Theta$ could be thought of as a topological charge.
The Gauss-Bonnet scalar of the internal space then reads $\hat G^{(4)}=4 \Theta+24 \kappa^2$ and the potential \cite{Dotti:2005rc}
\beq
V(r)=\kappa+\frac{r^2}{12 \alpha} \left(1\pm \sqrt{1+\frac{12}5\alpha\Lambda - 24 \frac{\alpha^2 \Theta}{r^4} + 24 \frac{\alpha M}{r^5}} \right)\, .
	\label{potentialclassIIEinstein}
\eeq
For $\Theta=0$, we obtain the well known black holes first discussed by Boulware and Deser (see \cite{Boulware:1985wk,Cai:2001dz}).

Alternatively (\ref{munueqclassII}) tells us that we can have a horizon which is potentially not Einstein, iff $B$ satisfies
\beq
1 + \alpha B^{ - 1/4} \left( {\frac{3}{2}\frac{{B'}}{B} - \frac{{B''}}{B'} + 8\frac{{BB'''}}{{B'^2 }} - 8\frac{{BB''^2 }}{{B'^3 }}} \right) = 0 \,.
	\label{munueqclassIInotEinstein}
\eeq
Note that in this case we have two equations for B and the system is overdetermined. Integrating \eqref{munueqclassIInotEinstein}, we obtain the following potential
\beq
	\tilde{V}(r) = \frac{r^2}{12\alpha}+\frac{\rho}{2\alpha}  - \frac{\mu}{2\alpha r},
\eeq
where $\mu$ and $\rho$ are integration constants. Comparing with \eqref{potentialclassII}, we make the following identifications :
\beq 5+12\alpha \Lambda=0, \quad \mu=0,\quad M=0.
\label{constclassIInotEinstein}
\eeq
and
\beq
\label{extra}
\rho=\frac{R^{\left( 4 \right)}}6\pm\frac16\sqrt{{R^{\left( 4 \right)}}^2-6\hat G^{\left( 4 \right)}}
\eeq
The potential \eqref{potentialclassII} reduces to
\beq
	V(r) =  \frac{\rho}{2}+ \frac{r^2}{12\alpha}\, .
\label{potentialclassIInotEinstein}
\eeq
This corresponds to a massless solution resembling adS or dS space, with a curvature  radius dependent on both the internal geometry and the Gauss-Bonnet coupling. The solution is defined only for $\left[R^{\left( 4 \right)}\right]^2-6\hat G^{\left( 4 \right)}>0$. Equation (\ref{extra}) is now a geometric equation constraining the 4-dimensional horizon geometry. Indeed $R^{\left( 4 \right)}$ and $G^{\left( 4 \right)}$ no longer have to be constant individually. In section \ref{sec:BIBS}, by Wick rotating these solutions to Lorentzian internal sections, we shall construct Born-Infeld black string solutions.

Thus, Class-II contains the folllowing solutions :
\begin{itemize}
	\item Class-IIa : The solution is locally static (\ref{Class2a}), and the horizon is an Einstein space with $\Theta\geq0$.
	\item Class-IIb : The solution is again locally static with potential given by (\ref{potentialclassIInotEinstein}), but the horizon is constrained by \eqref{extra} and the BI condition is imposed.
\end{itemize}

Thus, both subclasses of Class-II obey a local staticity theorem.

\subsection{Class-III}

The remaining Class of solutions is given by  $B=:\beta^4=$ constant $\neq 0$. In this case, the metric is no longer warped in the internal directions and the Einstein-Gauss-Bonnet equations (\ref{uv}), (\ref{munu}) reduce to
\begin{eqnarray}
0&=& -2 \Lambda \beta^4 + R^{(4)} \beta^2 + \alpha \hat{G}^{(4)} \label{beta} \\
G^{(4)}_{\mu \nu} + \Lambda \beta^2 h^{(4)}_{\mu\nu} &=& 2 \beta^3 \nu_{,uv} e^{-2\nu} \left ( \beta^2 h^{(4)}_{\mu\nu} - 4 \alpha G^{(4)}_{\mu\nu} \right ) \, .\label{NariaiGmunu}
\end{eqnarray}
It follows from contracting the second of the above equations, (\ref{NariaiGmunu}), with the metric $h^{ \mu\nu}$ that,
\beq 4\Lambda \beta^2 -R^{(4)} = 8 \beta^3 \nu_{,uv} e^{-2\nu} \left ( \beta^2 +  \alpha R^{(4)} \right )\, . \label{nunonsep}\eeq
If $R^{(4)}=-\beta^2/\alpha$, then we have the fine-tuning relation $1+ 4\Lambda \alpha =0$, (\ref{beta}) implies that $\hat{G}^{(4)} = \beta^4 /(2 \alpha^2)$ and (\ref{NariaiGmunu}) can be rewritten as
\beq \left (G^{(4)}_{\mu \nu} + \frac{1}{4} R^{(4)} h^{(4)}_{\mu\nu}\right ) \left ( \frac{2 \beta^3}{\Lambda} \nu_{,uv} e^{-2\nu} -1 \right )  = 0\, , \eeq
which implies that either $h^{(4)}_{\mu\nu}$ is Einstein and $\nu$ is not determined (and thus possibly time-dependent), or $h^{(4)}_{\mu\nu}$ is not necessarily Einstein and $\nu$ obeys the Liouville equation
\beq \nu_{,uv} = \frac{\Lambda}{2 \beta^3} e^{2\nu} \, .\eeq
The latter can be solved exactly, yielding
\beq e^{2\nu} = \frac{2\beta^3}{\Lambda} \, \frac{U' V'}{(U+V)^2} \, ,\eeq
for some functions $U=U(u)$ and $V=V(v)$. Now we can perform a change of coordinates of the form (\ref{uvUV}),
under which $\nu$ transforms in such a way that eventually
\beq
\label{z}
e^{2\nu} = \frac{2\beta^3}{\Lambda} \frac{1}{\bar{z}^2} \, .\eeq
The metric now obviously admits the locally time-like Killing vector $\partial_{\bar{t}}$ and Birkhoff's theorem holds in this case. Now, if on the contrary $R^{(4)}\neq -\beta^2/\alpha$, (\ref{nunonsep}) can be rewritten in the separable form
\beq \frac{4\Lambda \beta^2 -R^{(4)}}{\beta^2 +  \alpha R^{(4)}} = 8 \beta^3 \nu_{,uv} e^{-2\nu}  = \mbox{constant}\, . \label{nusep}\eeq
Provided that $1+ 4\Lambda \alpha \neq0$, we can have $R^{(4)} = 4\Lambda\beta^2$, which implies $\nu_{,uv}=0$ and $2\nu = \ln U' + \ln V'$ for some functions $U=U(u)$ and $V=V(v)$. Now we can perform a change of coordinates of the form (\ref{uvUV}) so that, in the end, $e^{2\nu}=1$ and the metric admits the Killing vector $\partial_{\bar{t}}$. It also follows from (\ref{beta}) that $\hat{G}^{(4)} = -2\Lambda \beta^4/\alpha$ and from (\ref{NariaiGmunu}) that $h^{(4)}_{\mu\nu}$ is Einstein. Otherwise, for non-vanishing values of the constant in (\ref{nusep}), say $\lambda$, $\nu$ obeys once again the Liouville equation
\beq \nu_{,uv} = \frac{\lambda}{8\beta^3} e^{2 \nu} \, .\eeq
After a change of coordinates of the form of (\ref{uvUV}), we therefore have
\beq e^{2\nu} = \frac{8\beta^3}{\lambda} \frac{1}{\bar{z}^2} \, ,\eeq
and the metric admits the Killing vector $\partial_{\bar{t}}$. If $\lambda=4\Lambda=-1/\alpha$, (\ref{NariaiGmunu}) is trivially satisfied and the only constraint on $h^{(4)}_{\mu\nu}$ comes from (\ref{beta}). Otherwise, it follows from (\ref{NariaiGmunu}) that $h^{(4)}_{\mu\nu}$ is Einstein and from (\ref{beta}) that $\hat{G}^{(4)}$ is a constant.

Wick rotating the solutions obtained in the former case, allows to construct axially symmetric black string type solutions, provided we impose a certain amount of symmetry to the internal manifold. Some static examples of this subclass of solutions have already been studied (see  \cite{Maeda:2006hj}, \cite{Maeda:2006iw} and references therein). We will briefly study an example in section \ref{sec:6dSBS}. It is worth noting that, once we allow for lesser symmetry, the scalar equation \eqref{beta} does not suffice to determine the full horizon metric.

The solutions contained in Class-III are the following :
\begin{itemize}
	\item Class-IIIa : $1+4\al\Lambda=0$, $R^{(4)}$, $\hat{G}^{(4)}$ are constant, and the horizon is Einstein.
	\item Class-IIIb : $1+4\al\Lambda\neq 0$, the transverse space is of constant curvature, and  \eqref{nusep} is satisfied, and the horizon is Einstein.
	\item Class-IIIc : $1+4\al\Lambda=0$, the transverse space is of constant curvature, and the horizon satisfies (\ref{beta}) and does not have to be Einstein.
\end{itemize}

Birkhoff's theorem holds for two of the subclasses, Class-IIIb and Class-IIIc.

\subsection{and a staticity theorem}
For generic Class-II and certain Class-III solutions, we have the following local staticity theorem.
\vskip.5cm

\noindent {\bf Theorem} {\it Let $({\mathcal M},g)$ be a six-dimensional pseudoriemannian spacetime whose metric $g$ satisfies the Gauss-Bonnet equations of motion (\ref{gbeq}) and whose manifold ${\mathcal M}$ admits a foliation into two-dimensional submanifolds $\Sigma_{(x_1, \dots x_4)}^{(2)}$ and a foliation
into four-dimensional submanifolds $H_{(t_1, t_2)} ^{(4)}$ such that :
\begin{itemize}
\item the tangent bundles of the leaves $T \Sigma_{(x_1, \dots, x_4)}^{(2)}$ and $T H_{(t_1, t_2)} ^{(4)}$ are orthogonal with respect to $g$;
\item for all $(t_1, t_2)$, the four-dimensional induced metric $h_{(t_1,t_2)}^{(4)}$ on $H_{(t_1, t_2)}^{(4)}$ is conformal to a given four dimensional metric $h^{(4)}$ with conformal factor depending only on $(t_1, t_2)$.
\end{itemize}
If in addition, either
\begin{enumerate}[i)]
	\item $1+4\Lambda \alpha \neq 0$ and $5+12 \alpha \Lambda \neq0$, or \label{casei}
	\item $1+4\Lambda \alpha = 0$ and $h^{(4)}$ is $\mathrm{not}$ an Einstein space, or \label{caseii}
	\item $5+12\al\Lambda=0$, $h^{(4)}$ is $\mathrm{not}$ an Einstein space and $R^{(4)}$ is not constant, \label{caseiii}
\end{enumerate}
then ${\mathcal M}$ admits a locally time-like Killing vector. Furthermore, in case \ref{casei}), $h^{(4)}$ is an Einstein metric with $\hat G^{(4)}= \mbox{constant}$, whereas in cases \ref{caseii}) and \ref{caseiii}), $h^{(4)}$ is not Einstein and solves respectively \eqref{beta} and \eqref{extra}.}
\vskip.5cm

\noindent This is a restatement of the properties of generic Class-II and some Class-III solutions we studied above, as these are the ones leading to necessarily static solutions. Note that the above theorem does not restrict the horizon geometry to be spherically symmetric. We can thus have horizons which are anisotropic as admissible static solutions. It should also be stressed that this is qualitatively different from the corresponding theorem in five dimensions, since there the black hole horizon is three-dimensional and its Weyl tensor is automatically zero. $D=6$ is the first case where the Weyl tensor $C_{\alpha \beta \gamma \delta}$ of the internal space plays a non-trivial role and can impose constraints. In dimensions $D>6$, we expect a similar situation, although one would be normally required to also consider the corresponding higher Lovelock densities in such a setup. The theorem of course makes no claims about the stability of such configurations. As we see, allowed horizons are four-dimensional Einstein spaces of Euclidean signature, with an added constraint on their Weyl tensor. Note that, since $\Theta$ is non-zero, in the non-compact cases these spaces are not asymptotically flat, for otherwise 
they should satisfy $C_{\alpha \beta \gamma \delta} \to 0$ at four-dimensional infinity.

\section{Horizon Structure}
We now focus on static Class-II solutions and elaborate on the form of the corresponding potential $V(r)$, (\ref{potentialclassIIEinstein}), which determines the occurrence of event horizons. In particular, we clarify the role of $\Theta$ in this case. There exists two branches of solutions, depending on the sign choice in (\ref{potentialclassIIEinstein}): the \emph{Einstein branch} solutions (-), which tend to Einstein solutions in the limit $\alpha\rightarrow 0$, and the \emph{Gauss-Bonnet branch} solutions (+), which have been argued to be unstable \cite{CharmousisPadilla2008}. Because of the stability problems associated with the latter, we restrict ourselves in the following on the Einstein branch, whose potential is given by
\beq
V\left( r \right) = \kappa  + \frac{{r^2 }}{{12\alpha }}\left( {1 - \sqrt {1 + \frac{12 \alpha \Lambda}{5}  - 24\Theta \frac{{\alpha ^2 }}{{r^4 }} + 24\alpha \frac{M}{{r^5 }}} } \right)   \,.
\label{bhpotential}
\eeq
In the following, we will then take $M$ to be positive, as is required to have a correct definition of mass in the usual $\Theta=0$ situation \cite{Deser:2002jk}. We should stress that once $\Theta\neq 0 $ the proper definition of  mass is no longer  clear, as the constant $\Theta$ changes the spacetime asymptotics. By continuity we take $M>0$, entrusting further study on the meaning of these charges to later work.

In the BI limit, $5+12\alpha \Lambda=0$, the only contributions come from the $\Theta$ and mass terms. At large $r$, the $\Theta\geq 0 $ term becomes dominant, developing a branch cut-type singularity. Solutions with $1+\frac{12 \alpha \Lambda}{5} =0$ and $\Theta\neq 0$ are therefore singular. The BI case thus falls into the second family of solutions verifying (\ref{lock}) which have to be treated separately.

From the above observation for the BI limit we already see that the $\Theta>0$ term will increase the possibility of a branch singularity near the BI limit. We assume for the rest of this section that $5+12\alpha \Lambda>0$.
A branch cut occurs at $r=r_\mathrm{bc}$ whenever
\beq
Q(r_\mathrm{bc}) = (1 + \frac{12 \alpha \Lambda}{5})r_\mathrm{bc}^5  - 24\Theta \alpha ^2 r_\mathrm{bc} + 24\alpha M=0\, .
\eeq
When does that actually happen? First, let us consider the simple case where $M$ is switched off. Then, provided $5+12\alpha\Lambda>0$, there is always a branch singularity at
\beq
	r_\mathrm{bc}=\left(\frac{24\alpha^2\Theta}{1+\frac{12\alpha\Lambda}5}\right)^{\frac14} =: 5^{1/4} r_0,
	\label{branchcut_M0}
\eeq
due to the non-vanishing of $\Theta$. On the other hand, if $M$ is not switched off, there is a branch-cut iff 
\beq
\alpha M <  \frac45\alpha^2\Theta r_0\, , \label{ineqM}
\eeq
where $r_0 >0$ is the minimum of $Q(r)$. The constraint (\ref{ineqM}) is the generalization of the $M=0$ result, the inequality on $M$ being trivially satisfied then. Generically, the effect of the $M$ term will be to 
decrease $r_\mathrm{bc}$, even if its exact expression cannot be computed analytically in the general case.

To go on, let us turn to the horizon analysis, first by considering the background solution, with $\Theta$ and $M$ switched off (or equivalently for $r$ large enough to make the $\Theta$ and $M$ terms negligible),
\beq
	V(r) = \frac {\left(1-\sqrt{1+\frac{12}{5}\alpha\Lambda}\right)}{12\alpha}\left(r^2-r_{\mathrm{c}}^2\right) = 0, \qquad r_c^2 = -\frac{12\alpha\kappa}{1-\sqrt{1+\frac{12}{5}\alpha\Lambda}},
\eeq
which is defined iff
\beq
	\kappa\Lambda>0, \quad \alpha\Lambda>-\frac5{12}.\\
	\label{condition_horizon_lambda}
\eeq
We obtain,
\bea
	V(r<r_c)>0 &\Longleftrightarrow& \Lambda>0, \nonumber\\
	V(r>r_c)>0 &\Longleftrightarrow& \Lambda<0, \nonumber
\eea
The solution behaves exacty like 4-dimensional AdS or dS space in GR with effective cosmological constant,
\beq
\Lambda_{eff}=\frac {\left(1-\sqrt{1+\frac{12}{5}\alpha\Lambda}\right)}{12\alpha}
\eeq

Now, as for the existence of event horizons, following \cite{Charmousis:2008kc} and \cite{Myers:1988ze}, $r=r_h$ is a horizon iff
\begin{itemize}
\item{$r_h>r_{bc}$}
\item{$r_h^2\geq -12\alpha \kappa$} (trivial if $\al\ka>0$)
\item{$r=r_h$ is a root of $P\left( r \right) = -\frac{\Lambda}{10} r^5  + \kappa r^3  + \alpha \left( {\Theta  + 6\kappa ^2 } \right)r - M$}
\end{itemize}
Whenever $\Theta=0$, the black holes behave similarly (modulo the branch singularity that puts some constraints on the smallness of the black hole mass) to their General Relativity black hole counterparts. Typically, $\Lambda<0$ permits planar and hyperbolic black holes, $\Lambda>0$ an event and a cosmological horizon, and $\Lambda=0$ a unique event horizon. The key question we want to answer here is: does $\Theta\neq 0$ introduce novel horizons to the above black holes, keeping in mind that $\Theta>0$? To answer this question, we momentarily switch off the ``mass" parameter $M$ and we note that if $\alpha<0$, the resulting black hole potential can be identified with that (tilded quantities) of the five dimensional Boulware and Deser solution \cite{Boulware:1985wk} (see also \cite{Cai:2001dz}), upon the following identifications
\beq
\tilde{\alpha}=3\alpha, \qquad \tilde \Lambda=\frac{3\Lambda}{5}, \qquad \Theta=\frac{-3\tilde M}{\tilde \alpha} \qquad M=0\, .
\eeq
Thus, we expect that horizons will be formed even if $M$ is set to zero.
In that case, $P(r)$ is a bisquare polynomial and its zeros $P(r_h>0)=0$ are easily found :
\beq
r_h^2=-\frac5\Lambda\left[-\kappa \pm \sqrt{\frac{2\al\Lambda}{5}\left(\Theta+\mathrm{sign}(\al\Lambda)\Theta_\mathrm{max}\right)}\right],
\eeq
where
\beq
	\frac{2\al\Lambda}5\left(\Theta+\mathrm{sign}(\al\Lambda)\Theta_{\mathrm{max}}\right)>0, \quad \Theta_{\mathrm{max}} = \frac{5\kappa^2}{2|\al\Lambda|}\left(1+\frac{12\al\Lambda}{5}\right).
\eeq
This inequality is always true if $\alpha\Lambda>0$, whereas when $\al\Lambda<0$ we need $\Theta<\Theta_\mathrm{max}$. These horizons, when defined, are always greater than the corresponding branch cut position $r_\mathrm{bc}$ \eqref{branchcut_M0}. When $\al\ka<0$, verifying $r_\mathrm{h}^2>-12\al\ka$ yields
\beq
	\Theta>\Theta_0, \quad \Theta_0=6\ka^2\left(1+\frac{12}5\al\Lambda\right).
\eeq
The occurrence of horizons due to the $\Theta$-term is summarized in the following Table \ref{TableM0}, for various signs of the cosmological constant and zero mass term. In short, $\Theta$ has no effect on the advent of horizons if $\al\ka>0$, whereas it will generate a new event horizon if $\al\ka<0$, for an infinite, bounded from below range of values when $\al\Lambda\geq0$ or for a finite range if $\al\Lambda<0$. It is quite interesting to see that there is a natural separation between these two cases, specifying clearly the effect of $\Theta$, depending on the respective signs of $\al\ka$.

\TABLE{
{\footnotesize
\begin{tabular}{|c||c|c||c|c||c|c|c|c|}
	\hline
	&\multicolumn{2}{|c||}{$\Lambda=0$}&\multicolumn{2}{|c||}{$\Lambda>0$ ($\ka>0$)}&\multicolumn{4}{|c|}{$\Lambda<0$}\\
	\hline
	 $\Theta$& $\al\ka>0$ & $\al\ka<0$&$\al >0$ & $\al<0$&$\ka<0$, $\al>0$&$\ka,\al>0$&$\ka>0$, $\al<0$&$\ka,\al<0$\\
	\hline
	 0 & $\varnothing$ &$\varnothing$&  \C &  \C& \K&$\varnothing$&$\varnothing$& \K\\
	\hline
	$\neq0$ & $\varnothing$ &  \E & \C& \C\,  + \E&\E+\K &$\varnothing$&\E  & \K \\
	&& iff $\Theta_0<\Theta$&&\multicolumn{2}{|c|}{iff $\Theta_0<\Theta<\Theta_{\textrm{max}}$ }&&iff $\Theta_0<\Theta$&\\
	\hline
\end{tabular}}
\caption{Occurrence of horizons, for parameter $M=0$, depending on the respective signs of $\kappa$ and $\alpha$. $\varnothing$ = no horizons, \E\, = Event horizon, \C\, = Cosmological horizon and \K\,= Killing horizon.  $\Theta_0=6\ka^2(1+\frac{12}5\al\Lambda)$, $\Theta_{\mathrm{max}}=\frac{5\Theta_0}{12|\al\Lambda|}$.}
\label{TableM0}
}

Let us now examine the special case of planar horizons ($\ka=0$) :
\begin{itemize}
	\item Usually, if $\Lambda=0$, no planar horizons are allowed. Here, there is one at $r_{\mathrm h}=\frac{M}{\al\Theta}$ provided $\al M>0$.
	\item For $\Lambda>0$, $M=0$, there is a cosmological horizon ($V(r>r_\mathrm{c})<0$) at $r_\mathrm{c}=10\frac{\al\Theta}\Lambda$ provided $\al>0$ (quite differently from the usual GR case).)
	\item For $\Lambda<0$, $M=0$, there is an event horizon ($V(r>r_\mathrm{h})>0$) at $r_\mathrm{h}=10\frac{(-\al)\Theta}{(-\Lambda)}$ provided $\al<0$.
\end{itemize}

If $M$ is not taken to be zero, it is difficult to evaluate quantitatively the impact of $\Theta$, and, apparently, little interesting information can be gained without resorting to a numerical study.

\section{Horizon Geometries in the Static Case}

After providing the general discussion of the theorem and the allowed static solutions, we proceed to give some concrete examples. As already mentioned, the geometry of the internal space on the horizon cannot be asymptotically flat due to the non-vanishing Weyl tensor. Candidate solutions are consequently not going to approximate flat space at infinity and we are led to consider geometries of this sort. Two simple examples of such configurations include an $S^{2} \times S^{2}$ geometry, as well as a variation of the Taub-NUT space, known as Bergman space. Finally, we will consider solutions that may have some interest for codimension two setups.

\subsection{$S^{2} \times S^{2}$}
This four-dimensional space is the product of two 2-spheres, with Euclidean signature and the metric
\beq
ds^2  = \rho _1 ^2 \left( {d\theta _1 ^2  + \sin ^2 \theta _1 d\phi _1 ^2 } \right) + \rho _2 ^2 \left( {d\theta _2 ^2  + \sin ^2 \theta _2 d\phi _2 ^2 } \right)\,,
\eeq
where we consider the (dimensionless) radii $\rho_{1}$ and $\rho_{2}$ of the spheres to be constant. The entire six-dimensional space has the form
\beq
ds^2  =  - V\left( r \right)dt^2  + \frac{{dr^2 }}{{V\left( r \right)}} + r^2 \rho _1 ^2 \left( {d\theta _1 ^2  + \sin ^2 \theta _1 d\phi _1 ^2 } \right) + r^2 \rho _2 ^2 \left( {d\theta _2 ^2  + \sin ^2 \theta _2 d\phi _2 ^2 } \right)\,,
\label{2sphere}
\eeq
with the potential
\beq
V\left( r \right) = \frac{{R^{\left( 4 \right)} }}{{12}} + \frac{{r^2 }}{{12\alpha}}\left( {1 \pm \sqrt {1 - 24k^2 \alpha - 24\Theta \frac{{\alpha ^2 }}{{r^4 }} + 24 \alpha \frac{M}{{r^5 }}} } \right)
\,.
\eeq
In order for (\ref{2sphere}) to be a solution to the Gauss-Bonnet equations of motion, we are led to the condition of equal sphere radii, $\rho_{1}=\rho_{2}$.
In that case, we have $\ka=\frac1{3\rho_1^2}>0$, $\Theta=\frac4{3\rho_1^4}$.
Since we want to look at the possible creation of an event horizon by
$\Theta$ if $M=0$, it suffices to check the case $\al<0$ for all values and
signs of the cosmological constant: Table \ref{TableM0} clearly shows that
such a creation only occurs as $\al\ka<0$, that is $\al<0$ in our case.
If $\Lambda=0$ or $\Lambda<0$, the constraint $\Theta_0<\Theta$ implies
\beq
        0\leq\al\Lambda<\frac5{12},
\eeq
which is trivially satisfied if $\Lambda=0$ and yields a \emph{minimum}
value for negative cosmological constant,
$\Lambda_{\textrm{min}}=\frac{5}{12\al}<0$.
On the other hand, if $\Lambda>0$, the constraint
$\Theta<\Theta_{\textrm{max}}$ (necessary to have any horizon at all)
implies
\beq
        -\frac5{36}<\al\Lambda<0,
\eeq
This gives this time a \emph{maximum} value for $\Lambda$,
$\Lambda_{\textrm{max}}=-\frac5{36\al}>0$, which more stringent constraint
than the one imposed to have a properly-defined background,
$5+12\al\Lambda>0$.

\subsection{Bergman Space}
The Bergman space is a homogenous but non-isotropic space which can be derived as a special case of the anti-deSitter Taub-NUT vacuum\cite{Chamblin:1998pz,Zoubos:2002cw}. The ordinary Taub-NUT metric\footnote{Since we consider the horizon geometry to carry a Euclidean signature, in this section all references to known metrics implicitly or explicitly assume a Euclidean version of them. These metrics are usually referred to in literature as gravitational instantons, since they represent solutions to Einstein's equations in Euclidean space with finite actions.} can be written as
\beq
\ud s^2  = W\left( \rho \right)\left( {\ud\tau ^2  + 2n\cos \theta \ud\phi } \right)^2  + \frac{{\ud\rho^2 }}{{W\left( \rho \right)}} + \left( {\rho^2  - n^2 } \right)\left( {\ud\theta ^2  + \sin ^2 \theta \ud\phi ^2 } \right) \,,
\label{TaubNUT1}
\eeq
with the potential $W(\rho)=\frac{\rho-n}{\rho+n}$. The Euclidean time coordinate has a period of $8\pi n$. Here, $n$ is what is usually called the ``nut'' parameter. It has dimensions of $mass^{-1}$. Mathematically, we define a nut as a zero-dimensional (point-like) space where the Killing vector generating the $U(1)$ Euclidean time isometry\footnote{The presence of this isometry is just a mathematical restatement of the property of the Taub-NUT solution being a static spacetime. In the case of Lorentzian Taub-NUT, the Killing vector shows the direction in spacetime (meaning, time $t$) towards which the metric remains unchanged. The isometry generated is thus a non-compact, one-parameter group of translations, while the parameter manifold is isomorphic to $R^{1}$. Once we Wick-rotate to imaginary time, $t \to i\tau$, Euclidean time $\tau$ becomes periodic and the parameter manifold is now $S^{1}$. The isometry, now generating rotations on the circle charactering the $\tau$ dimension turns into a $U(1)$.} vanishes. The nut is thus a fixed-point of the Euclidean time isometry. The Killing vector generating the isometry is in the case of Taub-NUT $K = \frac{\partial }{{\partial \tau }}$. A fixed-point occurs where $K=0$, or equivalently, $\left| K \right|^2  = g_{\mu \nu } K^\mu  K^\nu   = W\left( \rho \right) = 0$. Zeros of the Taub-NUT potential are then identified as positions of nuts. For the given potential, this occurs at $\rho=n$. We see that, at this position, the factor $\rho^{2}-n^{2}$ in front of the 2-sphere part of the metric is also zero, so the fixed-point set is really zero-dimensional as we would expect from the definition of a nut. This should be juxtaposed with the related concept of a ``bolt'', as a two-dimensional fixed-point {\it set}. We encounter such sets if the potential vanishes at some position different than $\rho=n$, which signifies the position of a two-dimensional sphere. In that sense, bolts are similar to black hole horizons, since they too are examples of such two-dimensional fixed-point sets for the Euclidean time isometry, although without a nut parameter. To have a regular solution for (\ref{TaubNUT1}), we only consider the range $\rho \ge n$.

In order to make contact with the parametrizations used for the description of the Bergman metric, we introduced the $SU(2)$ one-forms to parametrize the 3-sphere
\bean
\sigma _1  &=& \frac{1}{2}\left( {\cos \psi d\theta  + \sin \psi \sin \theta d\phi } \right) \,,\\
\sigma _2  &=& \frac{1}{2}\left( { - \sin \psi d\theta  + \cos \psi \sin \theta d\phi } \right) \,,\\
\sigma _3  &=& \frac{1}{2}\left( {d\psi  + \cos \theta d\phi } \right)\,.
\eean
These satisfy the cyclic relations $d\sigma _1  =  - 2\sigma _2  \wedge \sigma _3 $ etc. The angles $\theta, \phi, \psi$ vary in the ranges $0 \le \theta \le \pi$, $0 \le \phi \le 2\pi$, $0 \le \psi \le 4\pi$. The choice of parameters has to do with the asymptotic behavior of metric at infinity ($r \to 0$). There, the metric three remaining coordinates (angular and time) are combined to give a 3-sphere, which we parametrize using $\theta$, $\phi$ and $\psi$. We say that the metric is asymptotically locally flat, or ALF. This should be contrasted with the usual asymptotically flat (AF) metrics, where the corresponding boundary geometry at infinity is a direct product space $S^{1} \times S^{2}$, instead of $S^{3}$. For the Taub-NUT space, the time coordinate indices a non-trivial fibration of $S^{3}$.

Using the $SU(2)$ one-forms, and setting $\tau=2 n \psi$, we can eliminate the angular and time coordinates of the metric (\ref{TaubNUT1}) in favor of the one-forms. For the radial coordinate, we make the successive redefinitions $\rho \to \rho+n$, (so that $\rho$ starts at $\rho=0$) and then $\rho \to \frac{\rho^{2}}{2 n}$. The Taub-NUT metric can thus be rewritten as
\beq
\ud s^2  = 4\left( {1 - \mu ^2 \rho^2 } \right)\left[ {\ud\rho^2  + \rho^2 \left( {\sigma _1 ^2  + \sigma _2 ^2 } \right)} \right] + \frac{{4\rho^2 }}{{1 - \mu ^2 \rho^2 }}\sigma _3 ^2  \,,
\label{TaubNUT}
\eeq
where $\mu^{2}=\frac{1}{4n^{2}}$. The metric (\ref{TaubNUT}) can be considered to be a special case of the more general Anti-deSitter Taub-NUT, of the form
\beq
\ud s^2  = \frac{4}{{\left( {1 - k^2 \rho^2 } \right)^2 }}\left[ {\frac{{1 - \mu ^2 \rho^2 }}{{1 - k^2 \mu ^2 \rho^4 }}\ud\rho^2  + \rho^2 \left( {1 - \mu ^2 \rho^2 } \right)\left( {\sigma _1 ^2  + \sigma _2 ^2 } \right) + \rho^2 \frac{{1 - k^2 \mu ^2 \rho^4 }}{{1 - \mu ^2 \rho^2 }}\sigma _3 ^2 } \right]\,.
\eeq
Note that the mass parameter $\mu$ is now defined in terms of $k$ and the nut parameter by $\mu^{2}=k^{2}-\frac{1}{4n^{2}}$. This is a Taub-NUT space with a cosmological constant $-3k^{2}$. We consider the space of radial coordinates where the metric is non-singular, i.e. $0 \le \rho \le 1/k$, so that $\rho_{h}=1/k$ is the horizon of the $AdS$ space. For vanishing cosmological constant ($k=0$), this reduces to the ordinary Taub-NUT geometry of (\ref{TaubNUT}), while for $\mu=0$, the $AdS_{4}$ is recovered. $AdS$ Taub-NUT has in general an $SU(2) \times U(1)$ isometry group, which can however be enhanced for special parameter values.

None of the above mentioned spaces is a good candidate solution for the horizon, since they do not possess a constant $\Theta$. For $AdS$ Taub-NUT, we obtain
\beq
\Theta  = 6\mu ^4 \frac{{\left( {1 - k^2 \rho^2 } \right)^6 }}{{\left( {1 - \mu^{2} \rho^2 } \right)^6 }}\,,
\label{Theta}
\eeq
which only becomes constant at radial infinity (past the $AdS$ horizon), $\Theta  \sim \frac{{6k^{12} }}{{\mu ^8 }}$. Setting $k=0$ in this relation we obtain the corresponding value for the ordinary Taub-NUT, $\Theta=\frac{6\mu^{2}}{(1-\mu^{2} \rho^{2})^{6}}$. The space is asymptotically (locally) flat, so $\Theta \sim 0$ at infinity.

Let us now consider the case where $\mu=k$. We then recover the Bergman metric
\beq
\ud s^2  = \frac{4}{{\left( {1 - k^2 \rho^2 } \right)^2 }}\left[ {\frac{1}{{1 + k^2 \rho^2 }}\ud\rho^2  + \rho^2 \left( {1 - k^2 \rho^2 } \right)\left( {\sigma _1 ^2  + \sigma _2 ^2 } \right) + \rho^2 \left( {1 + k^2 \rho^2 } \right)\sigma _3 ^2 } \right]\,.
\label{Bergman}
\eeq
It describes the coset space $SU(2,1)/U(2)$, which is a K\"ahler-Einstein manifold with K\"ahler potential
\beq K(z_1, \bar{z}_1, z_2, \bar{z_2} ) = 1- z_1 \bar{z}_1 - z_2 \bar{z}_2 \, , \qquad \mbox{for $z_1 \bar{z}_1 + z_2 \bar{z}_2<1$,}\eeq
and the topology of the open ball in $\mathbb C^2$. Setting $z_1= k \xi \cos (\theta /2 ) e^{i(\phi+\psi)/2}$ and $z_2= k \xi \sin (\theta /2) e^{i(\phi-\psi)/2}$ the metric $g_{\alpha \bar{\beta}} = - \partial_\alpha \partial_{\bar{\beta}} \ln K^{1/k^2}$ reproduces exactly (\ref{Bergman}) after a change of coordinate $\xi^2=2 \rho^2/(1+k^2 \rho^2)$. The Bergman metric (\ref{Bergman}) has an isometry group of $SU(2,1)$. In practice, the choice $\mu=k$ corresponds to infinite ``squashing'' of the 3-sphere at the boundary $\rho \to 1/k$, such that only a one-dimensional circle remains intact at spatial infinity. By comparing the terms multiplying $\sigma_{1}^{2}+\sigma_{2}^{2}$ (2-sphere) and $\sigma_{3}^{2}$, we see that as we approach the boundary, the $\sigma_{3}^{2}$ part blows up faster and becomes dominant. The space has this circle as its conformal boundary. It is now possible to see from the expression (\ref{Theta}) for $\Theta$ in $AdS$ Taub-NUT that the Bergman space has $\Theta=6k^{4}$ and is thus a suitable horizon solution. Substituting (\ref{Bergman}) as the metric of the internal space $h^{(4)}_{\mu \nu}$, we verify that it is a solution to the equations of motion. To do so, we first rescale the radial coordinate as $\rho\to\rho/l$, with $l$ having dimensions of $mass^{-1}$ in order to make the metric dimensionless. As a result, we identify the dimensionless curvature scale $k\to kl$. The bulk potential of the solution is then given by
\beq
V\left( r \right) =  -k^{2}  + \frac{{r^2 }}{{12 \alpha}}\left( {1 \pm \sqrt {1 +\frac{12}5\alpha\Lambda - 144k^2 \frac{{\alpha ^2 }}{{r^4 }} + 24 \alpha\frac{M}{{r^5 }}} } \right)\,.
\eeq
Bergman space exists in the case $\ka=-k^2<0$,
$\Theta=6k^4$. According to Table \ref{TableM0}, when $M$ is set to zero,
the only case where a horizon may originate from the $\Theta$-term is when
$\al>0$ and $\Lambda$, the bulk cosmological constant, is negative. Then, the condition $\Theta_0<\Theta<\Theta_{\textrm{max}}$ needs
to be verified in order to have a new event horizon, on top of the
pre-existing Killing horizon.
The left part of the inequality yields $\al>0$ and is thus trivially
satisfied, and the right half gives a \emph{minimum} value for $\Lambda$,
\beq
        \Lambda_\textrm{min}=-\frac5{24\al}<\Lambda<0.
\eeq
This is a more stringent constraint than the one imposed to have a
properly-defined background, $5+12\al\Lambda>0$, which yields a lower
minimum value. If this is verified, the Bergman space with $M=0$,
$\Theta\neq0$ allows an event horizon.

We should note at this point that previous studies have shown the Bergman geometry to be unstable, both perturbatively and non-perturbatively,  in the context of ordinary General Relativity\cite{Kleban:2004bv}. It is not known whether this property persists also in Gauss-Bonnet theory.

As we mentioned above, apart from zero-dimensional fixed-points of the Euclidean time isometry (nuts), one could also consider spaces exhibiting the two-dimensional variety (bolts). This is known and appropriately termed as the Taub-Bolt space and is very similar to the already discussed Taub-NUT. Indeed, the metric for Taub-Bolt is the same as (\ref{TaubNUT1}) and (\ref{TaubNUT}), with the only distinction that the potential is now
\beq
W\left( \rho \right) = \frac{{\rho^2  - 2m\rho + n^2  + k^2 \left( {\rho^4  - 6n^2 \rho^2  - 3n^4 } \right)}}{{\rho^2  - n^2 }} \,.
\eeq
The position at which $W(\rho)=0$ is no longer $\rho=n$ and consequently the term $\rho^{2}-n^{2}$ multiplying the 2-sphere does not vanish at this point, providing the two-dimensional bolt. Imposing regularity of the potential at the position of the bolt $\rho=\rho_{b}$ we end up with the following prescriptions
\bea
m &=& \frac{{\rho_b ^2  + n^2 }}{{2\rho_b }} + \frac{{k^2 }}{2}\left( {\rho_b ^3  - 6n^2 \rho_b  - 3\frac{{n^4 }}{{\rho_b }}} \right) \\
\rho_{b \pm }  &=& \frac{1}{{12k^2 n}}\left( {1 \pm \sqrt {1 - 48k^2 n^2  + 144k^4 n^4 } } \right)
\eea
Is it possible to take the Bergman limit for the Taub-Bolt space like we did with Taub-NUT? To do so, we should retrace our steps and first recast the metric into the Pedersen form. Unfortunately, this is now non-trivial due to the more involved potential and bolt radius. We can however consider the limit $\mu=k$ without deriving the full metric for arbitrary $\mu$. Inspecting the definition of $\mu$ for Taub-NUT, we see that $\mu=k$ corresponds to the limit $n \to \infty$. To find the form of the metric in that limit, we first make the shift $\rho \to \rho+\rho_{b}$. The potential can then be written as
\beq
W\left( \rho \right) = \frac{{\rho\left( {C_0  + C_1 \rho + C_2 \rho^2  + C_3 \rho^3 } \right)}}{{(\rho + \rho_b  + n)(\rho + \rho_b  - n)}}
\eeq
with the parameters
\bea
C_0  &=& \frac{{\left( {\rho_b ^2  - n^2 } \right)\left( {1 + 3k^2 \left( {\rho_b ^2  - n^2 } \right)} \right)}}{{\rho_b }}\mathop  \sim \limits_{n \to \infty } 0 \,,\\
C_1  &=& 1 + 6k^2 \left( {\rho_b ^2  - n^2 } \right)\mathop  \sim \limits_{n \to \infty } 1 \,,\\
C_2  &=& 4k^2 \rho_b \mathop  \sim \limits_{n \to \infty } 4k^2 n \,,\\
C_3  &=& k^2 \,.
\eea
In determining the limit of parameters we used the fact that $\rho_{b} \mathop \sim \limits_{n \to \infty } n$. We then set $\rho \to \frac{\rho^{2}}{2n (1-k^{2}\rho^{2})}$ and keeping only finite terms in the metric, we recover the Bergman space (\ref{Bergman}). Taub-Bolt has thus the same limit as Taub-NUT for infinite nut parameter.

We would like to conclude this section by noting that, taking $k$ purely imaginary in (\ref{Bergman}), we end up with the Fubini-Study metric on $\mathbb{CP}^2$ and that the latter also constitutes a possible horizon metric for a static Lovelock black hole.

\subsection{Six-dimensional black strings}

Let us now turn to some special solutions which resemble black string metrics. Here we assume that the ``horizon" surface is of Lorentzian signature.
Both solutions presented in this section admit an extra axially symmetric Killing vector (see also \cite{Charmousis:2008bt}).

\subsubsection{Six-dimensional warped Born-Infeld black strings}
\label{sec:BIBS}
Throughout this section, the BI limit is assumed, that is we set $5+12\Lambda \alpha=0$. In this case, we would like to discuss a particular subclass of Class-II solutions, which appears to contain black string solutions as well as solutions that may be relevant to codimension two braneworld cosmology. They correspond to the overdetermined solutions (\ref{constclassIInotEinstein}-\ref{potentialclassIInotEinstein}). After Wick rotation, these solutions can be rewritten as
\beq ds^2 = r^2 h^{(4)}_{\mu\nu} dx^\mu dx^\nu + \frac{dr^2}{\frac{\rho}{2} + \frac{r^2}{12\alpha}} +\left(\frac{\rho}{2} + \frac{r^2}{12\alpha} \right ) d\theta^2 \eeq
where the four-dimensional Lorentzian metric $h^{(4)}_{\mu\nu}$ needs not be Einstein and is only subject to equation (\ref{extra}) that we reproduce here
\beq \rho=\frac{R^{\left( 4 \right)}}6\pm\frac16\sqrt{{R^{\left( 4 \right)}}^2-6\hat G^{\left( 4 \right)}} \, . \label{extrabis}\eeq
In order to solve (\ref{extrabis}), we assume, for example, that $h^{(4)}_{\mu\nu}$ is of the form
\beq
	\ud s^2_{\left( 4 \right)} = -f(\xi)\ud t^2 +\frac{\ud \xi^2}{f(\xi)} + \xi^2\ud \Omega_{II,k}^{2} \, , \label{sphansatz}
\eeq
where $\ud \Omega_{II,k}^{2}$ denotes the two-dimensional metric with constant curvature on the sphere, the plane or the hyperbolic space, depending on whether $k=1, 0$ or $-1$ respectively. $h^{(4)}_{\mu\nu}$ therefore has spherical, planar or hyperbolic symmetry, although it is certainly not the most general ansatz with these symmetries. Now, it follows from (\ref{extrabis}) that
\beq
	f(\xi)=k-\frac{\rho}2\xi^2\left(1\pm\sqrt{\frac{c_1}{\xi^3}+\frac{c_2}{\xi^4}}\right)\, ,
\eeq
where $c_1$ and $c_2$ are integration constants. The corresponding four dimensional metric $h^{(4)}_{\mu\nu}$ is not Einsein and distributional sources at $r^2= -6 \alpha \rho$ are therefore expected from the matching conditions.
These four dimensional metrics $h^{(4)}_{\mu\nu}$ do not correspond to any known GR solutions at large distance and are similar to the unphysical spherical solutions of Ho\v{r}ava gravity \cite{Horava:2009uw} in the case of detailed balance \cite{Lu:2009em}. Although BI and Ho\v{r}ava theory are radically different, both theories have been shown to suffer from strong coupling problems, \cite{Charmousis:2009tc}, \cite{CharmousisPadilla2008}.

The total space is, in the end, a warped product between a constant curvature two-space and a four-dimensional lorentzian space. This particular black string solution has been first discussed in \cite{CuadrosMelgar:2008kn}.

\subsubsection{Six-dimensional straight black strings}
\label{sec:6dSBS}
We finally consider the special case of Class-III solutions, with a time-like local Killing vector and an undetermined horizon geometry :
\beq
	\ud s^2 = \frac{2}{\Lambda \bar{z^2}}\left(-\ud t^2 + \ud z^2\right) + \beta^2h_{\mu\nu}\ud x^\mu\ud x^\nu.
\eeq
The only constraint on the internal geometry comes from the scalar equation \eqref{beta}, i.e.
\beq
	0=-2\Lambda\beta^4+\beta^2R^{\left( 4 \right)}+\al\hat G^{\left( 4 \right)}\, , \label{beta2}
\eeq
where $\beta$ is a constant ``warp factor" and $1+4\al\Lambda=0$. As in the previous section, we consider a Wick rotated version in which the internal space is lorentzian and we assume the same particular ansatz for $h^{(4)}_{\mu\nu}$, (\ref{sphansatz}). It then follows from (\ref{beta2}) that
\bea
	\ud s^2_{\left( 4 \right)} &=& -f(\rho)\ud t^2 +\frac{\ud \rho^2}{f(\rho)} +\rho^2\ud \Omega_{II}^{k} \\
	f(\rho)&=&k+\frac{\beta^2\rho^2}{4\al}\left[1\pm\sqrt{\frac{2}{3\beta^2}+\frac{32\al\mu}{3\beta^4\rho^3}-\frac{16\al q}{3\beta^4\rho^4}}\right],
\eea
where $\mu$ and $q$ are both integration constants. The have been rescaled so that the metric resembles the Reissner-Nordstr\"om solution far from the source in the minus branch, provided $\beta^2$ is set to two-thirds.

The six-dimensional metric finally reads
\beq
	\ud s^2 = \frac{2}{\Lambda z^2}\left(\ud\theta^2 + \ud z^2\right) + \beta^2\left[-f(\rho)\ud t^2 +\frac{\ud \rho^2}{f(\rho)} +\rho^2\ud \Omega_{II}^{k}\right]
\eeq
and is an unwarped product between a constant curvature two-dimensional space and a four-dimensional unwarped brane admitting Schwarzschild as a limit in one of the branches of solutions, with $\beta^2=\frac23$. This coincides with the Kaluza-Klein black hole reported in \cite{Maeda:2006hj}, provided $\beta^2=1$. We should emphasize here that, as an equation for $h^{(4)}_{\mu\nu}$, (\ref{beta2}) is underdetermined. In particular, had we considered a generic spherically symmetric ansatz, we would have had a free metric function appearing in the internal geometry.

\section{Conclusions}

We have found the general solution{\footnote{The case of Class(Ib) still demands the reoslution of (\ref{Beqn_class_notEinstein})}} to the metric  (\ref{metric})  and have investigated generalizations of Birkhoff's theorem in six-dimensional Einstein-Gauss-Bonnet theory (or Lovelock theory). Our analysis  significantly generalizes previous treatments in five dimensions and 6 dimensions, or cases where spherical symmetry of the horizon is imposed from the beginning. Furthermore, the analysis undertaken here agrees with \cite{Dotti:2008pp} where staticity is assumed. Permitting the Weyl tensor of the internal space in the equations of motion through the combination $C^{\alpha \beta \gamma \mu} C_{\alpha \beta \gamma \nu}=\Theta \delta^{\mu}_{\nu}$ leads to severe restrictions. We analyzed the way this new contribution modifies the available solutions. We distinguish three categories.

The so called Class-I leads both to an underdetermined system of equations and the application of a specific condition between the parameters of the theory. We find two possibilities :
\begin{itemize}
\item the internal space is a constant curvature space (with $\Theta=0$) and one of the metric functions in transverse space is undetermined (Ia),
\item the internal space is not necessarily Einstein (and generically $\Theta\neq0$) and all metric functions can be determined (Ib).
\end{itemize}

The possibility of an underdetermined system of equations once a particular choice of parameters is used seems to hint the presence of an increased ``symmetry'' in such a case. Class-I solutions do not obey some variant of Birkhoff's theorem, i.e. static solutions are not unique in this context. Class-II solutions on the other hand give rise to a generalized Birkhoff's theorem; static solutions are unique, provided some conditions related to the structure of the internal space are satisfied :
\begin{itemize}
\item the internal space is Einstein with a constant 4-dimensional Gauss-Bonnet charge  and constant curvature (IIa),
\item the internal space is not necessarily Einstein but is constrained by a scalar equation \eqref{extra} and the BI condition holds (IIb).
\end{itemize}

The Class-III case corresponds to unwarped metrics, and Birkhoff's theorem also holds in some specific subcases :
\begin{itemize}
\item  $1+4\al\Lambda\neq0$ and the internal space is Einstein (IIIb), or
\item $1+4\al\Lambda=0$, the internal space is not Einstein and can or not be constrained by a scalar equation \eqref{beta} (IIIc).
\end{itemize}
A third case exists where Birkhoff's theorem does not hold, when both the horizon is Einstein and the condition $1+4\al\Lambda=0$ is applied (IIIa).

We summarize our results in Table \ref{TableResults}.

\TABLE{
\centering
\begin{tabular}{|c|c|c||c|c||c|c|c|}
	\hline
		& Ia & Ib & IIa & IIb & IIIa & IIIb & IIIc \\
	\hline
	\bf{Birkhoff} & $\varnothing$& $\varnothing$ & $\surd$ & $\surd$ & $\varnothing$& $\surd$& $\surd$\\
	\hline
	\bf{Einstein} & $\surd $& $\varnothing $ &$\surd$ & $\varnothing$, \eqref{extra}&$\surd $ &$\surd$ &$\varnothing$, \eqref{beta} \\
	\hline	
	$\Theta$ & $0$ & $\geq 0$ & $\geq 0$ & $\geq 0$ & $>0$ & $\geq0$ & $\geq0$ \\
	\hline
	\bf{Fine-tuning} & \bf{BI}& \bf{BI}& $\varnothing$&  \bf{BI}& $1+4\al\Lambda=0$ & $\varnothing$&$1+4\al\Lambda=0$ \\
	\hline
\end{tabular}
\caption{Classes of solutions and their characteristics. {\bf Einstein} : horizon is an Einstein space. {\bf BI} : $5+12\al\Lambda=0$. $\Theta\doteq\frac14C^{abcd}C_{abcd}$.}
\label{TableResults}
}

For the Class-II solutions, for which the generalized staticity theorem holds, we studied some examples of non-trivial horizon geometries. The spaces we consider are in general anisotropic, such as the $S^{2}\times S^{2}$ product space and the Euclidean Bergman geometry. The latter can be considered as the appropriate limit of either an AdS Taub-NUT or Taub-Bolt space with infinite nut charge. Bergman space has the squashed 3-sphere (Berger sphere) as its conformal boundary and is thus anisotropic.

It would be interesting to investigate further cases of suitable horizon geometries satisfying the requirements of Birkhoff's theorem and also to study the general conditions under which a class of such solutions may arise. A consistent generalization to higher dimensions would require the inclusion of higher order Lovelock densities in the action. In this case one could consider as possible candidate horizon solutions the Bohm metrics \cite{Gibbons:2002th}, which are known to be admissible if only the Gauss-Bonnet term is taken into account. Apparently, higher-order curvature invariants other than $\Theta$ would be involved in distinguishing compatible horizon metrics, potentially requiring a more systematic classification.

The most interesting departure from General Relativity arises due to the non-vanishing of the constant $\Theta$. The latter appears, at the level of the static black hole potential, as a novel integration constant or ``charge'' and
is directly related to the Gauss-Bonnet scalar of the 4-dimensional horizon, a quantity whose integral yields a topological invariant: the relevant Euler-Poincar\'e characteristic.
We saw that the presence of this constant imposes particular and non-trivial asymptotic conditions and certainly a particular topology. Since it can even give rise to novel horizons, it would be interesting to investigate whether this constant can be interpreted as the conserved charge of some Killing symmetry of spacetime and what its physical meaning actually is.

\section*{Acknowledgements}

We would like to thank B. Boisseau,  R. Emparan, B. Linet, N. D. Kaloper, G. Niz, T. H. Padilla, P. Saffin, S. Solodukhin and K. Zoubos for interesting discussions and comments on the subject. CB
acknowledges support from the CNRS and the Universit\'e de Paris-Sud XI. RZ is grateful to the EPSRC for financial support. CC thanks his colleagues in the Yukawa Institute for their hospitality and discussions on the subject of Lovelock theory during his stay in Kyoto at the initial stages of this work.

\bibliographystyle{JHEP}
\bibliography{references}

\end{document}